# Scaling behavior in steady-state contractile actomyosin network flow

**Maya Malik-Garbi[1], Niv Ierushalmi[1], Silvia Jansen[2,3], Enas Abu-Shah[1,4], Bruce L. Goode[2], Alex Mogilner[5] and Kinneret Keren[1,6]**

[1] Department of Physics, Technion- Israel Institute of Technology, Haifa 32000, Israel

[2] Department of Biology, Brandeis University, Waltham, MA, USA

[3] Department of Cell Biology and Physiology, Washington University St. Louis, St. Louis, MO, USA

[4] Kennedy Institute of Rheumatology, University of Oxford, Oxford OX3 7FY, UK

[5] Courant Institute of Mathematical Sciences and Department of Biology, New York University, New York, NY 10012, USA

[6] Network Biology Research Laboratories and Russell Berrie Nanotechnology Institute, Technion – Israel Institute of Technology, Haifa 32000, Israel

**Contractile actomyosin network flows are crucial for many cellular processes including cell division and motility, morphogenesis and transport. How local remodeling of actin architecture tunes stress production and dissipation and regulates large-scale network flow remains poorly understood. Here, we generate contractile actomyosin networks with rapid turnover *in vitro*, by encapsulating cytoplasmic *Xenopus* egg extracts into cell-sized 'water-in-oil' droplets. Within minutes, the networks reach a dynamic steady-state with continuous inward flow. The networks exhibit homogenous, density-independent contraction for a wide range of physiological conditions, indicating that the myosin-generated stress driving contraction is proportional to the effective network viscosity. We further find that the contraction rate approximately scales with the network turnover rate, but this relation breaks down in the presence of excessive crosslinking or branching. Our findings suggest that cells use diverse biochemical mechanisms to generate robust, yet tunable, actin flows by regulating two parameters: turnover rate and network geometry.**

The dynamic organization of cellular actin networks emerges from the collective activities of a host of actin-associated proteins, including factors that stimulate actin assembly and disassembly, various crosslinkers and filament binding proteins which define the local network architecture, and myosin motors that generate contractile forces [1-3]. Myosin activity can drive global contraction and generate contractile actin network flows, which play a crucial role in cell division, polarization and movement. For example, cell division is driven by the contraction of the cytokinetic ring, which is a quasi-1D contractile actomyosin network [4]. Contractile flows in the actin cortex are essential for cell polarization [5], while bulk actin network flows are crucial for localization of cellular components during the early stages of embryo development [6]. Continuous retrograde actin network flows were further shown to be the basis for amoeboid cell motility, the primary mode of motility observed *in vivo* for cells moving in a confined 3D geometry [7, 8]. While it appears from all these examples that rapid actin turnover is necessary to sustain continuous stress generation and drive persistent network flow, there is limited quantitative understanding of what determines the contractile behavior of these networks, and in particular what governs the rate of network flow. Both the force generation within actomyosin networks and their viscoelastic properties depend in non-trivial ways on the architecture and turnover dynamics of the network [9-15]. Conversely, the network architecture and dynamics are influenced by the stress distribution in the network and the flows it generates [15-18]. As such, understanding how the microscopic dynamics of actin and its associated proteins shape the large-scale structure and flow of contractile networks remains an outstanding challenge.

The study of contractile actin networks has primarily focused on the interplay between myosin motor activity and network connectivity [12, 16, 19]. *In vitro* reconstitution allows considerable control of the mechano-chemical conditions, which is difficult to achieve *in vivo*. Indeed, experiments with reconstituted networks assembled from purified actin, crosslinkers and myosin motors established the conditions required for large-scale global contraction [10, 16, 20-24]. However, these *in vitro* actin networks that were reconstituted using purified proteins displayed negligible rates of disassembly, which are orders of magnitude slower than actin turnover rates *in vivo* [3]. In the absence of rapid turnover, myosin-motor activity led to transient and essentially irreversible contraction; this is very different from the persistent actin flows in living cells, which undergo continuous recycling of network components. Recent theoretical modeling and simulation studies have started to explore how rapid turnover influences force production, dissipation and the dynamic spatio-temporal organization of contractile actomyosin networks [13, 14, 25-27]. In particular, these simulations suggest that rapid turnover is essential for generating persistent flows, allowing the network to continuously produce active stress and dissipate elastic stress, while maintaining its structural integrity [13].

Experimental efforts to study contractile actin networks undergoing rapid turnover have been hampered by the lack of a suitable model system. *In vitro* reconstituted contractile networks typically have limited turnover, since actin disassembly in pure solutions is slow in the absence of additional disassembly factors [3], whereas studies on contractile networks *in vivo* are limited by the difficulties of measuring actin dynamics in complex cell geometries and the ability to control and vary system parameters. Here we leverage the unique benefits of an *in vitro* reconstitution system based on cell extracts that provides both myosin activity and physiological actin turnover rates, to investigate how the interplay between actin assembly and disassembly, myosin motor activity and network connectivity, determine large-scale structure and dynamics of contractile actin networks. Importantly, the rapid turnover in our system allows the reconstituted networks to attain a dynamic *steady-state* characterized by persistent contractile network flow and a self-organized, radially symmetric density distribution.

Using this *in vitro* system we are able to quantitatively characterize actin network turnover and flow, and systematically investigate the large-scale emerging properties of contractile actin networks. We find that under a broad range of physiological conditions, the networks contract homogenously, despite large spatial variations in network density. The emergence of a density-independent contraction rate is surprising since the internal force generation in actin networks and their viscoelastic properties are strongly dependent on network density [11, 28, 29]. Theoretical modeling using 'active gel theory' suggests that the homogenous, density-independent contraction arises from a scaling relation between the active stress and the effective viscosity in these self-organized networks. We further find that the contraction rate is roughly proportional to the network turnover rate as we vary the dynamics of the system by changing its composition. This scaling relation breaks down with excessive branching or crosslinking. Together, these results show how physiological rates of actin turnover influence force production and dissipation in cellular actomyosin networks, demonstrating that the rate of network flow depends both on the actin turnover rate and on the local network architecture.

**Results and Discussion**

To generate contractile actin networks, we encapsulate cytoplasmic *Xenopus* egg extract in water-in-oil emulsion droplets [30-32]. Endogenous actin nucleation activities induce the formation of a bulk network, which undergoes myosin-driven contraction [16, 33]. Within minutes after droplet formation, the system assumes a dynamic steady-state characterized by an inward network flow and a stationary network density that decreases toward the periphery (Figs. 1, S1; Movie 1). This steady-state is made possible by the rapid actin turnover in the system, such that the inward network flux is balanced by a diffusive flux of disassembled network components back to the periphery. The inward network velocity approaches zero at the boundary of the dense spherical 'exclusion zone' which forms at the center of the droplet. This exclusion zone appears within minutes, as the network contracts and accumulates particulates from the (crude) extract, condensing this material into a spheroid (Fig. 1b, S1e).

The simple geometry and persistent dynamics of this system facilitate quantitative analysis of actin network turnover and flow. The network is labeled using a low concentration of GFP-Lifeact, which does not significantly alter the network (see Methods), and imaged by time-lapse spinning disk confocal microscopy. Images are acquired at the mid-plane of the droplets, where the network flow is approximately planar due to the symmetry of the set up. The network density and flow have a spherically-symmetric pattern (Fig. 1b,c), which remains at steady-state for more than half an hour (Fig. S1).

Interestingly, the inward flow velocity, $V$, increases nearly linearly as a function of distance from the inner boundary, $V \approx -k(r - r_0)$, with $k$ a constant, $r$ the radial coordinate, and $r_0$ the radius of the exclusion zone (Fig. 1e). A linear velocity profile with $V \sim -kr$ is a signature of uniform global contraction; in a homogenously contracting network the relative velocity between two points is directly proportional to their distance, so the radial network flow velocity depends linearly on the radial distance from the stationary contraction center (see Supplementary Information). The contraction rate in this case can be determined from the slope of the linear fit for the inward radial velocity as a function of distance, and is equal to $k=0.65 \pm 0.15$ min$^{-1}$ (mean±std; $N$=39). This constant rate of contraction is a global characteristic of the self-organized network dynamics, and is not dependent on the geometry of the droplet (Fig. S2).

In general, the relationship between the network contraction rate and the network density is non–trivial, since the active stress driving contraction and the viscoelastic properties of the network are both density-dependent [11, 28, 29]. Our observations here, that the self-organized dynamics in the system lead to a local contraction rate that is uniform across the system, despite large spatial variations in network density, is thus surprising. As shown below, homogenous network contraction is observed for a range of different physiological conditions, suggesting that the density-independent contraction rate is a manifestation of an inherent scaling relation in the system, rather than the result of fine-tuning of parameters.

We can infer the net rate of actin turnover taking into account the conservation of actin subunits described by the continuity equation, $\frac{\partial \rho}{\partial t} + \nabla \cdot \vec{J} =$ net turnover, where $\rho$ is the network density, $\vec{V}$ is the network velocity, $\vec{J} = \rho \vec{V}$ is the network flux (Fig. 1f), and the net turnover quantifies the local changes to network density due to actin assembly and disassembly events. The network density and velocity fields in our system are at steady-state (Fig. S1), so the net turnover (i.e. the difference between local assembly and disassembly) will be equal to the divergence of the network flux. Using the measured network density and flow (Fig. 1d,e), we can thus determine the spatial distribution of network turnover ($\nabla \cdot \vec{J}(r)$; Fig. 1g). We find that the network undergoes net disassembly (i.e. $\nabla \cdot \vec{J} < 0$) closer to the contraction center in regions with higher network density, and net assembly (i.e. $\nabla \cdot \vec{J} > 0$) toward the periphery where the network is sparse. The net turnover can also be plotted as a function of the local network density, showing that disassembly increases with network density (Fig. 1h). We can approximate the relation between network turnover and density by a linear fit: $\nabla \cdot \vec{J}(\rho) \sim \alpha - \beta \rho$ (Fig. 1h). This linear relation describes the simplest model for actin turnover, with a constant assembly term and disassembly that is proportional to local network density. The net turnover rate, $\beta$, can be estimated from the slope of the linear fit to the divergence of network flux as a function of network density (Fig. 1h; see Methods), which is equal to $\beta$= 1.4 ±0.3 min$^{-1}$ ($N$=39).

The network dynamics can be modified by supplementing the extract with various components of the actin machinery. Actin disassembly can be enhanced by the combined action of the severing/depolymerizing protein Cofilin together with Coronin and Aip1, which have recently been shown to work in concert to enhance actin disassembly *in vitro* even under assembly-promoting conditions (e.g., physiological conditions with high actin monomer concentrations as present in our system) [34]. Our results show that addition of these three proteins (CCA: Cofilin, Coronin, and AIP1) together in the reconstituted system (Fig. 2a; Movie 2) induces a ~1.5-fold increase in the net actin turnover ($\beta$= 2.1 ±0.3 min$^{-1}$, $N$=7). Addition of Cofilin, Coronin, or Aip1 individually does not enhance turnover (Fig. S3), illustrating that their combined activity is required for promoting disassembly under physiological conditions. The presence of CCA also leads to faster network contraction (Fig. 2a). Under these conditions the network still contracts homogenously, with a linear velocity profile, but at an increased contraction rate of $k$= 1.3 ±0.15 min$^{-1}$ ($N$=7). Similar effects are observed when actin assembly is restricted with Capping Protein that caps free barbed ends (Fig. S4), whereas stabilizing actin filaments by adding phalloidin has the opposite effect (Fig. S5a).

Actin assembly can be enhanced by adding nucleation-promoting factors (Fig. 2c-g; Movie 3). Adding increasing amounts of ActA, which activates the Arp2/3 complex to nucleate branched actin filaments, has a dramatic effect on actin network distribution and flow. The rate of contraction gradually decreases with increasing concentrations of ActA, until at high concentrations (1.5µM) the contraction nearly stops (Fig. 2b,d,e). The slower contraction also leads to an increase in the extent of the network, so that at high ActA concentrations the network nearly fills the entire droplet (Fig. 2b). The net turnover rate remains largely unchanged (Fig. S6a).

Supplementing the extract with mDia1, which nucleates unbranched actin filaments, has a qualitatively different effect (Fig. 2c,g). The dependence of the contraction rate on the concentration of mDia1 becomes non-monotonic: adding mDia1 up to a concentration of 0.5 µM leads to a small decrease in the contraction rate, but further addition of the nucleator reverses the trend and the contraction rate increases. At high mDia1 concentrations (~1.5 µM), the contraction rate returns to values close to the control sample, but the network appearance is very different, with a strong diffuse background signal, likely due to the presence of many filaments which are not associated with the network. Again, the net turnover rate remains largely unchanged (Fig. S6b).

The qualitative difference between the two types of nucleators is likely related to the inherent difference in the connectivity of the nucleated filaments. Arp2/3 mediated nucleation creates a new actin filament that is physically connected to an existing filament through a branch junction. In contrast, mDia1 nucleates new filaments in solution which must subsequently be crosslinked to other filaments to become associated with the network. As such, high levels of Arp2/3 activation generate highly connected networks consisting of densely branched filaments, whereas enhanced mDia1 activity results in many unconnected actin filaments.

Our results (Figs. 1-2) demonstrate that confined contractile actin networks with rapid turnover can attain a dynamic steady-state over a wide range of parameters. Achieving a steady-state does not require fine-tuning of system parameters; rather the system dynamically adapts the overall assembly/disassembly rates to be equal (e.g. by varying the fraction of polymerized actin). Our results shows that the net actin disassembly increases with network density at a roughly constant rate (Fig. 1h). While we cannot measure actin assembly and disassembly separately here, these findings suggest that the actin turnover rate is relatively insensitive to the myosin-induced stress. In this scenario, the actin disassembly rate is roughly fixed by the turnover machinery. The system reaches a steady state by tuning the actin monomer concentration – so both the fraction of actin associated with the network and the assembly rate change for the system to reach steady state (and will depend on the disassembly rate). The characteristics of the steady-states obtained vary considerably as a function of system composition, whereby the rate of network contraction and/or the turnover rate can shift over a few-fold range by adding different components of the actin machinery (Figs. 2, S3-S5). Surprisingly, the network flow maintains a nearly linear velocity profile under a variety of different conditions. Thus, while the rate of network contraction depends on the composition of the system, the local contraction rate for a given condition is homogenous, across a range of network densities.

To understand the origin of this homogenous contractile behavior we turned to modeling using active gel theory [35, 36]. The network is described as an active isotropic actomyosin gel characterized by a density $\rho$ and a velocity field *V*. The network dynamics are governed by two equations, one for mass conservation (continuity equation) and one for momentum conservation (force balance equation). At steady-state the force balance equation can be written as $\nabla \cdot \sigma = \nabla \cdot (\sigma_A + \sigma_{VE}) = f_D$, where $\sigma_A$ is the active stress, $\sigma_{VE}$ is the effective viscoelastic network stress, and $f_D$ is the friction with the surrounding fluid. The viscoelastic contribution is dominated by the viscous stress (Supp. Information), and estimates indicate that the friction with the fluid is negligible (Supp. Information). Hence, the force balance equation can be described as a balance between the myosin-generated active stress which drives contraction, and the viscous network stress resisting it: $\nabla \cdot (\sigma_A + \sigma_V) \approx 0$. Assuming that the actomyosin gel is an isotropic compressible viscous continuum in a spherically-symmetric system, the force-balance equation has the following form, $\frac{\partial}{\partial r}\left(\sigma_A + 2\mu \frac{\partial V}{\partial r} + \lambda \nabla \cdot V + 4\mu \frac{V}{r}\right) = 0$, where $\lambda$ and $\mu$ are the bulk and shear network viscosities, respectively [37]. The periphery of the network at large radii typically does not reach the water-oil interface and the friction with the cytosol is negligible, so we take stress free boundary conditions for the outer rim of the network. Integrating over *r* and taking into account this stress-free boundary condition at large *r*, we obtain: $\sigma_A + \frac{(\lambda + 2\mu)}{r^2} \frac{\partial}{\partial r}(r^2 V) = 0$. The local contraction rate (and hence the velocity profile) in this case depends only on the ratio between the myosin-generated active stress and the effective viscosity of the network, $-\frac{1}{r^2} \frac{\partial}{\partial r}(r^2 V) = \frac{\sigma_A}{2\mu + \lambda}$. Both the active stress and the network viscosities are expected to vary considerably, since the network density is spatially variable. However, if the ratio between them remains constant, $\frac{\sigma_A(r)}{2\mu(r) + \lambda(r)} = c$, the velocity profile will be linear $\frac{1}{r^2} \frac{\partial}{\partial r}(r^2 V) = -c \rightarrow V = -\frac{c}{3} r$, and the network will contract homogenously with a contraction rate equal to $k = c/3$. This analysis indicates that when the active stress and the effective viscosities scale similarly with density, the local contraction rate becomes density-independent and the network contracts homogenously, despite large spatial density variations, as observed (Figs.1-2). The most likely scenario is that both the active stress and the network viscosities increase linearly with network density, so their ratio remains constant. Based on our theoretical analysis, we can use the measured velocity profile to deduce the ratio between the active stress and the effective viscosity, $\frac{\sigma_A}{2\mu+\lambda}$, which is found to be of order $\sim 1\,\text{min}^{-1}$ for the different conditions examined.

This analysis also indicates that the observation of a non-linear contraction velocity profile implies that the ratio between the active stress and the effective viscosity varies in space and is hence density-dependent. We find such behavior when we supplement the extract with sufficiently high concentration of crosslinkers such as α-Actinin (Fig. 3; Movie 4). We can infer the density-dependence of the ratio between the active stress and the viscosity from the spatial variation in the local contraction velocity (Fig. 3e). We find that the ratio between the active stress and the viscosity, for high α-Actinin concentration (>4µM), decreases with density, indicating that non-linear contributions as a function of density in the active stress, the effective viscosity, or both, become important. Addition of the filament bundler Fascin had a different effect; Actin network contraction remained homogenous but at a reduced rate (Fig. S5b), and the net turnover rate was lower. The difference between α-Actinin and Fascin is likely related to their different actin binding properties, suggesting that the addition of α-Actinin increases network viscosity in a non-linear fashion.

The addition of different actin regulators (e.g. nucleation promoting factors, disassembly factors, bundlers or crosslinking proteins) at concentrations comparable to their physiological concentration [38], allowed us to explore contractile network behavior over a wide range of physiologically-relevant conditions (Fig. 4). The network dynamics are characterized by two time scales, (1) the contraction time ($k^{-1}$), and (2) the actin turnover time ($\beta^{-1}$), which have the same order of magnitude (~1 min). The diffusion time across the system provides a third timescale ($\tau_D \sim R^2/6D$, where $R$ is the droplet radius and $D \sim 15\,\mu m^2/s$ is the diffusion coefficient of actin monomers in the cytosol [39]). However, for droplets in the size range considered here (R<65 µm), the diffusion time is < 1min. In this case, diffusion is sufficient to redistribute network components across the system, and the behavior of the system is primarily governed by the contraction and turnover time scales.

We consider the characteristic time scales for network dynamics over the range of physiological conditions examined (Fig. 4). The contraction rate and the turnover rate are determined from the slopes of the linear fits of the velocity as a function of distance ($k$) and the net turnover as a function of density ($β$), respectively (Fig. 4a,b). For many conditions (e.g. addition of CCA, Fascin, mDia1, Capping Protein), the network contraction rate roughly scales with the net turnover rate (dashed line; Fig. 4c), such that their ratio remains nearly constant. As a result, the density profiles of the contracting networks are similar, and the main difference between different conditions is the faster or slower dynamics (Fig. 4d). Simulation studies show that in the presence of rapid turnover, viscous dissipation in the network will be dominated by filament breaking and network turnover, so the effective viscosity is expected to be inversely proportional to the turnover rate [13, 40]. This is consistent with our results, since according to the model (see above) the contraction rate should be inversely proportional to the effective viscosity. Thus, if the viscosity is inversely proportional to the turnover rate, the contraction rate will scale with the turnover rate as observed. Adding crosslinkers (α-Actinin) or enhancing branching activity (ActA) leads to large deviations in the ratio between the turnover rate and contraction rate, which are reflected by large-scale changes in network structure (Fig. 4); Faster contraction results in a compact network concentrated around the contraction center as observed with α-Actinin, whereas slower contraction results in a more extended network. The local modifications in network architecture generated by increasing the density of filament branch junctions or crosslinks, are thus translated to global changes in the overall network structure.

The behavior of contractile actomyosin networks arises from the interplay between the active stress generated within the network and the viscous stress of the network. Both the force generation and the dissipation depend in a non-trivial manner on the microscopic details of the network organization which are determined by the activity of a host of actin regulatory proteins, including assembly and disassembly promoting factors, various crosslinkers and actin binding proteins as well as myosin motors. We have developed an experimental system where we can for the first time systematically investigate the emergent behavior of contractile networks at steady-state and quantitatively characterize network flow and turnover in the presence of physiological rates of actin turnover. Our findings suggest that the active stress and the effective network viscosity scale linearly with network density, so their ratio becomes density-independent, leading to homogenous network contraction. We further show that the contraction rate scales with the actin turnover rate and depends on the local network architecture. These observations provide new insights into the relationship between network contraction and density *in vivo*, which has been studied e.g. in the context of the cytokinetic ring [4]. Our results suggest that a constant ring constriction rate can arise, even if the network density changes over time, if the production and dissipation of internal stress in the cytokinetic ring scale with the density as observed here. Furthermore, the rate of constriction would decrease when actin disassembly processes are slowed down, as observed experimentally [41]. More generally, the ability to study contractile networks with physiological actin turnover rates *in vitro,* as demonstrated here, can provide further insight into the behavior of actomyosin networks *in vivo* and promote quantitative understanding of cellular dynamics.

## Materials and Methods

*Cell extracts, Proteins and Reagents*

Concentrated M-phase extracts were prepared from freshly laid *Xenopus laevis* eggs as previously described [31-33]. Briefly, *Xenopus* frogs were injected with hormones to induce ovulation and laying of unfertilized eggs for extract preparation. The eggs from different frogs were pooled together and washed with 1× MMR (100 mM NaCl, 2 mM KCl, 1mM $MgCl_2$, 2 mM $CaCl_2$, 0.1 mM EDTA, 5 mM Hepes, pH 7.8, at 16°C. The jelly envelope surrounding the eggs was dissolved using 2% cysteine solution (in 100 mM KCl, 2 mM $MgCl_2$, and 0.1 mM $CaCl_2$, pH 7.8). Finally, eggs were washed with CSF-XB 10 mM K-Hepes pH 7.7, 100 mM KCl, 1 mM $MgCl_2$, 5 mM EGTA, 0.1 mM $CaCl_2$, and 50 mM sucrose containing protease inhibitors (10 µg/ml each of leupeptin, pepstatin and chymostatin). The eggs were then packed using a clinical centrifuge and crushed by centrifugation at 15000g for 15 minutes at 4 °C. The crude extract (the middle yellowish layer out of three layers) was collected, supplemented with 50 mM sucrose containing protease inhibitors (10 µg/ml each of leupeptin, pepstatin and chymostatin), flash frozen as 10 µl aliquots and stored at -80 °C. Typically, for each extract batch a few hundred aliquots were made. Different extract batches exhibit similar behavior qualitatively, but the values of the contraction rate and disassembly rate vary (Fig. S7). All comparative analysis between conditions was done using the same batch of extract.

ActA-His was purified from strain JAT084 of *L. monocytogenes* (a gift from Julie Theriot, Stanford University) expressing a truncated actA gene encoding amino acids 1–613 with a COOH-terminal six-histidine tag replacing the transmembrane domain, as described in [31, 32]. Purified proteins were aliquoted, snap-frozen in liquid $N_2$, and stored at -80°C until use.

Cor1B and AIP1 were expressed and purified from transfected HEK293T cells (ATCC). Cells were grown on plates at 37°C under a humidified atmosphere containing 5% $CO_2$ in Dulbecco's modified Eagle's medium (DMEM), supplemented with 10% (v/v) heat-inactivated foetal bovine serum, glucose (4.5 g/l), penicillin (100 units/ml) and streptomycin (100 µg/ml). Cells at 30-40% confluence were transiently transfected using 25 kDa linear polyethylenimine (Polysciences, Warrington, PA). 72 h post-transfection, cells were harvested in PBS, pelleted by centrifugation at $1,000 \times g$ for 5 min, and lysed by repeated freeze-thawing in 20 mM Tris/HCl pH 7.5, 150 mM NaCl, 1% (v/v) Triton X-100 and a standard cocktail of protease inhibitors (Roche, Germany). After a 30 min incubation on ice, cell lysates were cleared by centrifugation at 20,000 × $g$ at 4°C using an eppendorf tabletop centrifuge and incubated with $Ni^{2+}$-NTA beads (Qiagen, Valencia, CA) for 90 min at 4°C in the presence of 20 mM imidazole. After washing with Buffer A (20 mM Tris pH 7.5, 300 mM NaCl, 50 mM imidazole and 1 mM DTT), proteins were eluted in Buffer A supplemented with 250 mM imidazole, concentrated, and purified further on a Superose 6 gel filtration column (GE Healthcare Biosciences, Pittsburgh, PA) equilibrated in Buffer B (20 mM Tris pH 8.0, 50 mM KCl and 1 mM DTT). Purified proteins were concentrated, aliquoted, snap-frozen in liquid $N_2$, and stored at -80°C until use.

Human Cofilin-1 (Cof1) was expressed in BL21 (DE3) *E. coli* by growing cells at 37°C in TB medium to log phase, then inducing expression with 1 mM isopropyl β-D-1-thiogalactopyranoside (IPTG) at 18°C for 16 h. Cells were harvested by centrifugation and stored at -80°C, then lysed by sonication in 20 mM Tris pH 8.0, 50 mM NaCl, 1 mM DTT and protease inhibitors. Lysates were cleared by centrifugation at 30,000 × $g$ for 20 min in a Fiberlite F13-14X50CY rotor (Thermo Scientific, Rockport, Illinois), and applied to a 5 ml HiTrap HP Q column (GE Healthcare Biosciences). The flow-through containing Cof1 was collected and dialyzed into 20 mM Hepes pH 6.8, 25 mM NaCl and 1 mM DTT. Next, the protein was applied to a 5 ml HiTrap SP FF column (GE Healthcare Biosciences) and eluted with a linear gradient of NaCl (25 to 500 mM). Fractions containing Cof1 were concentrated and dialyzed into Buffer B, aliquoted, snap-frozen in liquid $N_2$, and stored at -80°C until use.

α-Actinin was purchased from Cytoskeleton Inc., and reconstituted to final concentration of 40µM with water. Fascin was purchased from Prospec, dialyzed against Hepes pH 7.5 and reconstituted to final concentration of 20µM in XB buffer with 50mM Hepes.

Actin networks were labeled with GFP-Lifeact (gift from Chris Fields). The purified protein was concentrated to a final concertation of 252µM in 100mM KCL, 1mM MgCl2, 0.1mM CaCl2, 1mM DTT and 10% Sucrose.

*Emulsion preparation*

An aqueous mix was prepared by mixing the following: 8µL crude extract, 0.5 µL 20× ATP regenerating mix (150mM creatine phosphate, 20mM ATP, 20mM MgCl2 and 20mM EGTA) 0.5 µM GFP-Lifeact and any additional proteins as indicated. The final volume was adjusted to 10 µL by adding XB (10 mM Hepes, 5 mM EGTA, 100 mM KCl, 2 mM $MgCl_2$, 0.1 mM $CaCl_2$ at pH 7.8). The concentration of the components of the actin machinery in the mix can be estimated based on [38]. The total actin concentration is estimated to be ~20 µM. The ATP regeneration mix enables the system to continuously flow for more than 1-2 hours.

Emulsions were made by adding 1-3% (v/v) extract mix to degassed mineral oil (Sigma) containing 4% Cetyl PEG/PPG-10/1 Dimethiocone (Abil EM90, Evnok Industries) and stirring for 1 min on ice. The mix was then incubated for an additional 10 min on ice to allow the emulsions to settle. Samples were made in chambers assembled from two passivated coverslips separated by 30 µm-thick double stick tape (3M), sealed with vaseline:lanolin:paraffin (at 1:1:1) and attached to a glass slide. Passivation is done by incubating cleaned coverslips in silanization solution (5% dichlorodimethylsilane in heptane) for 20 minutes, washing in heptane, sonicating twice in DDW for 5 minutes and once in ethanol for 5 minutes, and drying in an oven at 100 ºC. Droplets with radiis in the range of 25-65 µm were imaged 10-60 min after sample preparation.

*Microscopy*

Emulsions were imaged on a 3I spinning disk confocal microscope running Slidebook software, using a 63× oil objective (NA=1.4). Images were acquired using 488nm laser illumination and appropriate emission filters at room temperature. Images were collected on the spinning disk confocal with an EM-CCD (QuantEM; Photometrix). Time lapse movies of emulsions were taken at the equatorial plane, so the network velocity is within the imaging plane.

*Analysis*

The steady-state density and velocity distributions were determined from time lapse movies of contracting networks. The movies were taken at time intervals of 2.5-10 sec (depending on network speed) with 512x512 pixels per frame at 0.2054 micron/pixel. Background subtraction and corrections for uneven illumination field were done by subtracting the mean intensity of images of droplets without a fluorescent probe, and dividing by a normalized image of the illumination field distribution. To determine the illumination field distribution, we imaged very large droplets (larger than the field of view) with uniform probe distribution generated by inhibiting actin polymerization with 6.6 µM Capping Protein and no ATP regeneration mix.

Bleach correction was done by dividing the entire image by a constant factor determined from an exponential fit to the total image intensity as a function of time. In addition, the differences in the refractive index of the oil and the extract within the droplets, distorts the intensity near the edge of the droplets. To correct for this, we measured the fluorescent intensity as a function of distance from the edge for droplets with uniform probe distribution (as above). The correction as a function of distance from the edge of the droplet was determined by averaging the measured intensity near the edge for ~50 droplets. The fluorescent signal intensity was corrected for edge effects by dividing the intensity by this correction function.

The velocity distribution (Fig. 1c) was extracted using direct cross-correlation analysis based on PIVlab code [42] written in matlab with modifications. The movies were first preprocessed to enhance contrast using contrast limited adaptive histogram equalization (CLAHE) and high pass filter in Matlab. Cross-correlation was done on overlapping regions 30×30 pixels in size, on a grid with 10 pixel intervals. To avoid artefacts associated with padding the boundaries in the correlation analysis, the 30×30 correlation analysis was done using 60×60 windows from the original images and no padding was used. The correlation functions from pairs of consecutive images from the movie was averaged over time (between 5-40 consecutive image pairs). The peak of the time-averaged correlation function was determined by fitting a 2D Gaussian, and the local velocity in each correlation region was determined from the shift of the Gaussian's peak from the origin. To automatically exclude spurious correlations, we considered only regions in which the fit was to a Gaussian with a single peak that is more than 3 pixels away from the window boundary, has a peak amplitude that is larger than 5-20% (depending on the conditions used) of the average correlation peak value, and has a width (at 70% height) which is less than 20 pixels. To further remove spurious correlations at large radii (where the network density is close to background levels, and hence the correlation signal is weak) we excluded velocity vectors whose angular component was larger than a threshold. The radial velocity was determined by averaging the velocities as a function of distance from the contraction center, for radii for which the velocity was determined for more than half the relevant grid points. The radius of the inner boundary ($r_0$) was determined as the intersection point of the linear fit to the velocity with the x-axis. The contraction rate for each droplet was determined from the slope of the linear fit to the radial velocity as a function of distance from the contraction center.

The network density was determined by averaging the corrected fluorescence signal over different angles and over time (typically 20-40 frames) to obtain the probe density distribution as a function of distance from the contraction center. The measured probe distribution reflects the sum of the signal emanating from the network bound probe and from the diffusing probe. To approximate the network bound probe density we subtracted the measured signal near the periphery of the droplet, which is equivalent to assuming that the network density approaches zero near the periphery and that the diffusing probe is distributed nearly uniformly.

Analysis of net turnover of the network is based on the continuity equation at steady-state, $\nabla \cdot \vec{J} = \text{net turnover}$. The network flux is the product of the network density and velocity, $\vec{J}(r) = J(r)\hat{r} = \rho(r)V(r)\hat{r}$. The divergence of the flux for the spherically-symmetric case is equal to $\nabla \cdot \vec{J}(r) = \frac{1}{r^2}\frac{d(r^2 \rho(r) V(r))}{dr}$. The divergence of the flux was also plotted as a function of the measured density distribution $\rho$, and the disassembly rate was determined from the slope of the best fit linear approximation $\nabla \cdot \vec{J}(\rho) \approx \alpha - \beta\rho$.

**Acknowledgements**

We thank Yariv Kafri, Erez Braun and members of our lab for useful discussions and comments on the manuscript. We thank Nikta Fahkri, Tzer Han Tan, Alex Solon, Raphael Voituriez, Guy Bunin, and Christoph Schmidt for useful discussions. We thank Chris Field and James Pelletier for advice on extracts and for the GFP-Lifeact construct. We thank Gidi Ben-Yoseph for excellent technical support. We thank Julian Eskin for assistance with cartoon models.

This work was supported by a grant from the Israel Science Foundation (grant No. 957/15) to K.K., a grant from the United States-Israel Binational Science Foundation (grant No. 2013275) to K.K. and A.M, by US Army Research Office grant W911NF-17-1-0417 to A.M., and by grants from the Brandeis NSF MRSEC DMR-1420382 and the National Institutes of Health (R01-GM063691) to B.G.

# Figures

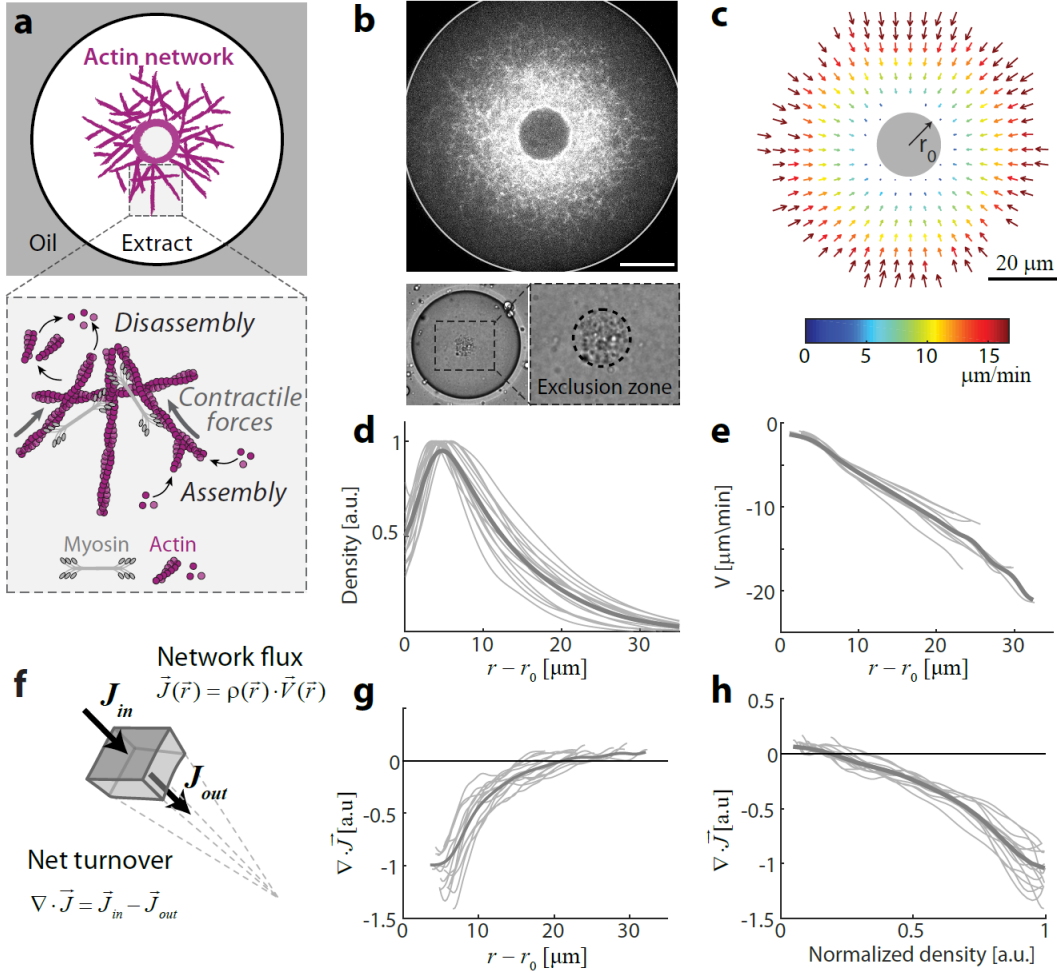

**Figure 1. Quantitative analysis of actin network flow and turnover.** Bulk contractile actin networks are formed in 'artificial cells' made from *Xenopus* extract encapsulated within water-in-oil droplets. The system reaches a dynamic steady-state characterized by a spherically-symmetric inward flow. (a) Schematic of the actin network inside a water-in-oil droplet, illustrating the actin turnover dynamics and myosin-driven contraction. (b) Top: Spinning disk confocal image of the equatorial cross section of a network labeled with GFP-Lifeact. Bottom: Bright-field images of a droplet showing the aggregate of particulates that forms the exclusion zone around the contraction center. (c) The network velocity field for the droplet shown in (b), as determined by correlation analysis of the time-lapse movie (Movie 1). (d) The actin network density as a function of distance from the contraction center. The thin lines depict data from individual droplets, and the thick line is the average density profile. The density is normalized to have a peak intensity =1. (e) The radial

velocity as a function of distance from the contraction center. The inward velocity increases linearly with distance. (f-h) Analysis of net actin network turnover. (f) Schematic illustration showing that at steady-state, the divergence of the actin network flux is equal to the net network turnover rate. The network flux is equal to the product of the local network density and velocity ***J(r)*** = ρ(***r***)***V(r)***. (g) The divergence of the network flux, $\nabla \cdot \vec{J}(r)$, is plotted as a function of distance from the contraction center, showing the spatial distribution of the net turnover. Negative values (at smaller *r* values) correspond to net disassembly, while positive values (near the droplet's periphery) indicate net assembly. (h) The divergence of the network flux is plotted as a function of the local actin network density, $\nabla \cdot \vec{J}(\rho)$. The net turnover decreases roughly linearly with actin network density.

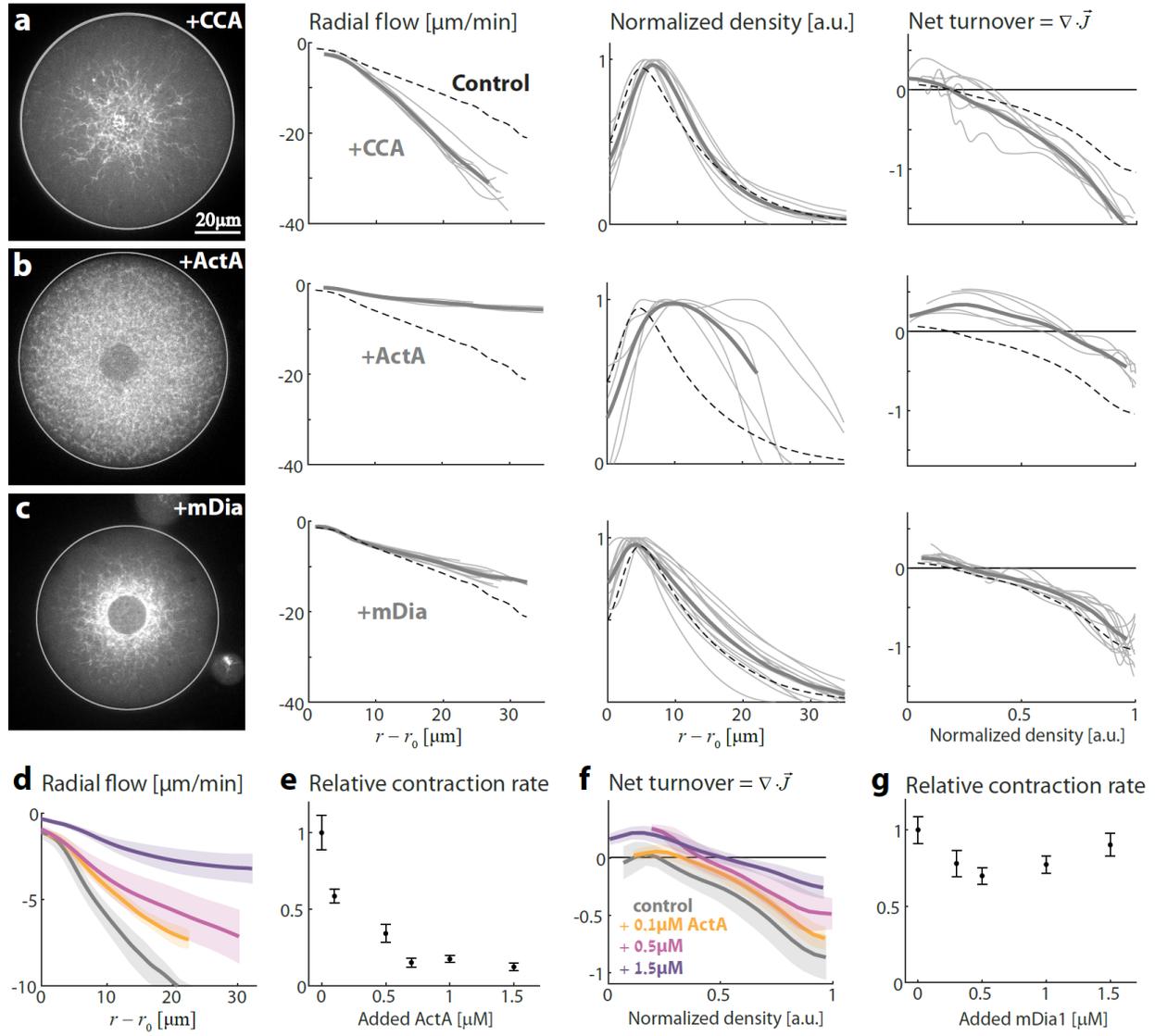

**Figure 2. Influence of assembly and disassembly factors on actin network architecture, flow and turnover.** Contractile actin networks are generated in 'artificial cells' by encapsulating *Xenopus* extract supplemented with different assembly and disassembly factors. In all cases, the system reaches a steady-state within minutes. The inward contractile flow and actin network density were measured (as in Fig. 1). (a-c) The steady-state network behavior is shown for samples supplemented with (a) 12.5µM Cofilin, 1.3µM Coronin and 1.3 µM Aip1 (see Movie 2); (b) 1.5µM ActA (see Movie 3); (c) 0.5µM mDia1. For each condition, a spinning disk confocal fluorescence image of the equatorial cross section of the network labeled with GFP-Lifeact (left) is shown, together with graphs depicting the radial network flow and density as a function of distance from the contraction center (middle), and the net actin turnover as a function of network density (right).

The thin grey lines depict data from individual droplets, and the thick line is the average over different droplets. The dashed lines show the results for the control unsupplemented sample. (d-g) The concentration-dependent effect of adding ActA (d-f) and mDia1 (g) on network dynamics. (d) The radial network flow is plotted as a function of distance from the contraction center. For each ActA concentration, the mean (line) and standard deviation (shaded region) over different droplets are depicted. (e) The network contraction rate, determined from the slope of the radial network flow as a function of distance (shown in d), is plotted as a function of the added ActA concentration. The contraction rate decreases with increasing ActA concentration. (f) The net actin turnover rate, determined from the divergence of the flux, is plotted as a function of network density for the different ActA concentrations. (g) The network contraction rate, determined from the slope of the radial network flow as a function of distance, is plotted as a function of the added mDia1 concentration.

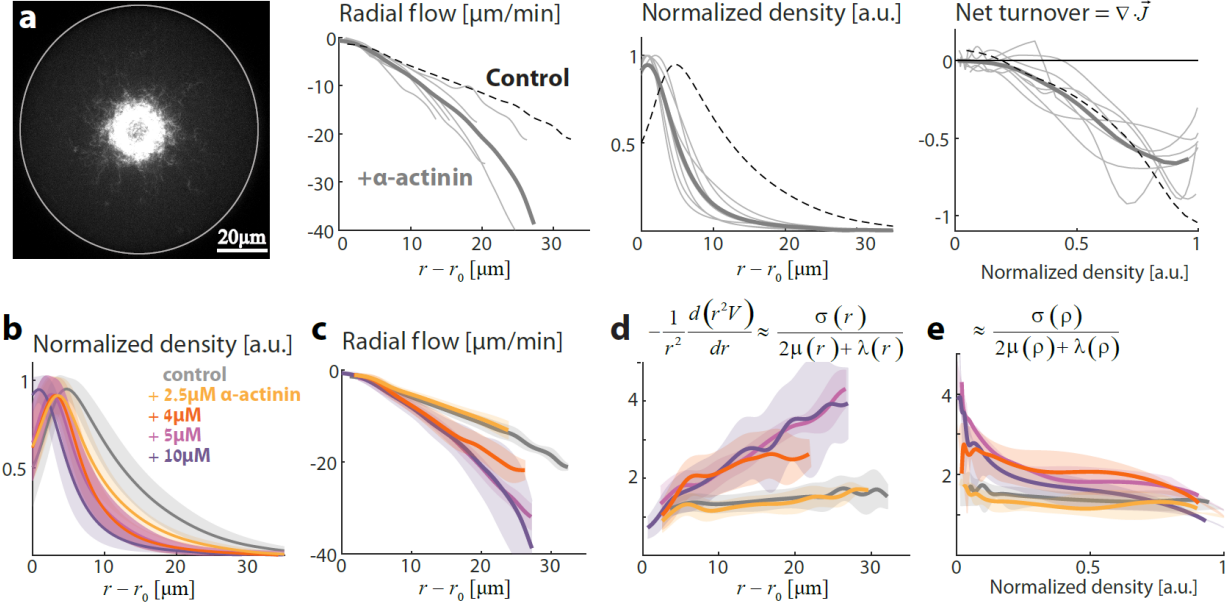

**Figure 3. Influence of excessive crosslinking on actin network dynamics.** Contractile actin networks are generated in 'artificial cells' by encapsulating *Xenopus* extract supplemented with different concentrations of the actin crosslinker α-Actinin. The inward contractile flow and actin network density were measured (as in Fig. 1). (a) The steady-state network behavior is shown for a sample supplemented with 10 μM α-Actinin (see Movie 4). A spinning disk confocal fluorescence image of the equatorial cross section of the network labeled with GFP-Lifeact (left) is shown, together with graphs depicting the inward radial network flow and network density as a function of distance from the contraction center (middle) and the net actin turnover as a function of network density (right). The thin grey lines depict data from individual droplets, and the thick line is the average over different droplets. The dashed lines show the results for the control unsupplemented sample. The network contracts in a non-homogenous manner, reflected by the non-linear dependence of the radial network flow on the distance from the contraction center. (b,c) The concentration-dependent effect of α-Actinin on network density and flow. The network density (b) and radial flow (c) are plotted as a function of distance from the contraction center. For each α-Actinin concentration, the mean (line) and standard deviation (shaded region) over different droplets are depicted. The position of the peak network density moves towards the inner boundary with increasing α-Actinin concentrations, and the radial velocity becomes a non-linear function of the distance from the contraction center. (d) The derivative of the radial velocity, $-\frac{1}{r^2}\frac{\partial}{\partial r}(r^2 V)$, is plotted as a function of distance from the contraction center. This function becomes position-

dependent for α-Actinin concentrations $\geq 4\,\mu M$. According to the model, this derivative should be approximately equal to the ratio between the active stress and the effective network viscosities, $\dfrac{\sigma_A(r)}{2\mu(r)+\lambda(r)}$. (e) The derivative of the radial velocity, $-\dfrac{1}{r^2}\dfrac{\partial}{\partial r}(r^2 V)$, is plotted as a function of network density. According to the model, this derivative should be approximately equal to the ratio between the active stress and the effective network viscosities, $\dfrac{\sigma_A(\rho)}{2\mu(\rho)+\lambda(\rho)}$. For α-Actinin concentrations $\geq 4\,\mu M$ this ratio becomes density-dependent, indicating that the scaling relation between the active stress and the effective viscosity no longer holds.

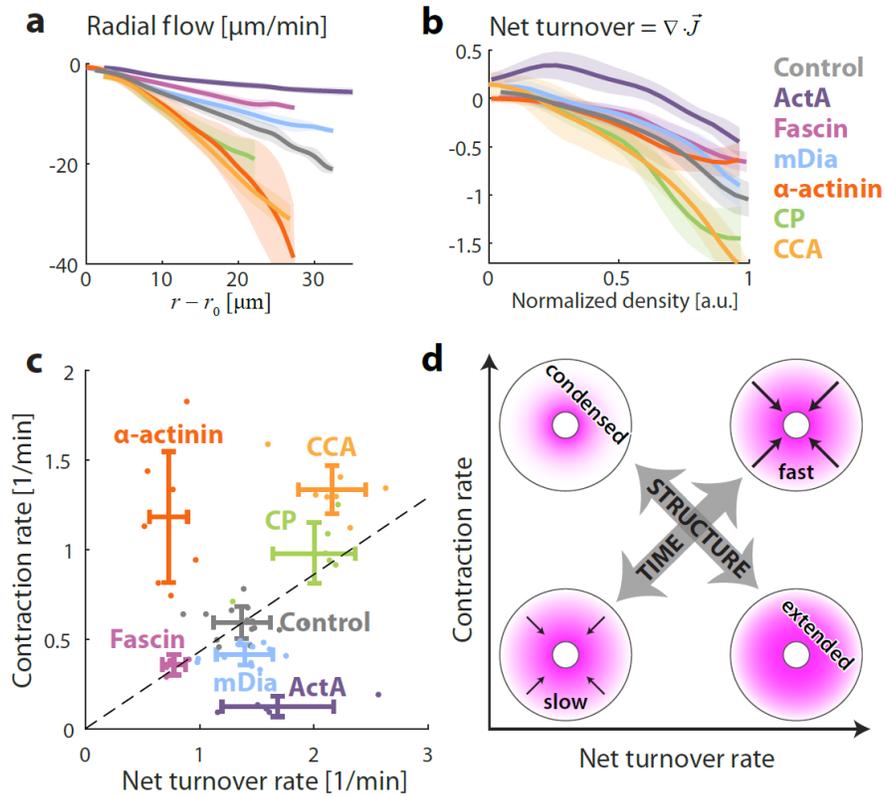

**Figure 4. Characteristics of contractile actin networks with turnover.** The contractile behavior and actin turnover were measured for contractile actin networks formed in 'artificial cells' under different conditions as indicated. (a) The radial network flow rate is plotted as a function of distance from the contraction center. Data is shown for 80% extract (control) and for samples supplemented with 1.5μM ActA; 2.6μM Fascin; 0.5μM mDia1; 10μM α-Actinin; 0.5μM Capping Protein; or 12.5μM Cofilin, 1.3μM Coronin and 1.3 μM Aip1. For each condition, the mean (line) and standard deviation (shaded region) over different droplets are depicted. (b) The net actin turnover as a function of network density is plotted for different conditions. (c) A scatter plot of the net contraction rate and net turnover rate for the different conditions tested. The contraction rate and turnover rate were determined for each droplet from the slopes of linear fits to the radial network flow as a function of distance, and the net turnover as a function of density, respectively. For each condition, the dots depict values for individual droplets and the error bars show the mean and standard deviation for all the droplets examined for the given condition. The dashed line indicates where the ratio between the contraction rate and turnover rate is equal to the control sample. (d) Scheme showing how the behavior of the network changes as a function of the contraction and turnover rates. Increasing the contraction and turnover rates proportionally leads

to faster network dynamics but no obvious change in network structure, whereas changes in the ratio between the contraction and turnover rates leads to significant modifications in network structure.

# Supplementary Information

# Scaling behavior in steady-state contractile actomyosin network flow

Maya Malik-Garbi, Niv Ierushalmi, Silvia Jansen, Enas Abu-Shah, Bruce L. Goode, Alex Mogilner and Kinneret Keren

## Supplementary Text

**Viscoelastic contribution is dominated by the viscous stress**

Generally, cellular actin networks are viscoelastic, with complex mechanical properties that can be approximated by a combination of Maxwell and Kelvin-Voight models [1-5]. These cited studies show that if a continuous stress is applied to an actin network, there is a short-term elastic response followed by a long-term viscous, flow-like behavior. The viscoelastic relaxation time – the time after which the elastic response fades and the viscous behavior ensues – was measured to be as short as a few seconds [3-5] (or even ~ 0.1 second [1]), much shorter than the characteristic timescales of actin network turnover and contraction, ~ 1 minute, in our system. Moreover, the elastic part of the viscoelastic stress and non-Newtonian components of the stress in the actin networks on long time scales were shown to be less than 10% of the total stress [6]. Therefore, we model the actin network as a purely viscous fluid.

**Friction between the fluid and actin network is negligible**

The Darcy friction forces between the porous polymer mesh and the fluid squeezing through it can be estimated as the characteristic actin network flow rate, ~ 0.1 µm/s, divided by the hydraulic permeability of the cytoskeleton, ~ 0.01 µm$^3$/(pN×s) [7], so the order of magnitude of the respective stress is ~ 10 pN/µm$^2$. The characteristic myosin contractile stress is usually on the order of ~ 100 pN/µm$^2$ [8, 9]. Measurements of the characteristic viscoelastic network stress that balances the myosin contractile stress also gave estimates of ~ 100 pN/µm$^2$ [10]. Thus, the Darcy friction forces between the actin mesh and the fluid squeezing through it are an order magnitude smaller than the myosin contractile stress and the viscoelastic network stress, and hence can be neglected.

**Network contraction with an excluded region**

We consider a spherically-symmetric network with a radial flow velocity field $V(r)$ and an inner exclusion zone of radius $r_0$. For a homogenously contracting network, the divergence of the velocity field will be constant and equal to the volume contraction rate, $c$, so that,

$div(V) = \frac{1}{r^2}\frac{\partial}{\partial r}(r^2 V) = -c$ and $V(r_0) = 0$. The velocity profile in this case will be

$V(r) = -\frac{c}{3}r\left(1 - \frac{r_0^3}{r^3}\right)$. For $r_0 = 0$, we obtain a linear velocity profile $V(r) = -kr$ with $k = \frac{c}{3}$. For $r_0 > 0$, the velocity profile has a non-linear correction and the slope of the velocity profile is equal to $\frac{dV}{dr} = -\frac{c}{3}\left(1 + 2\frac{r_0^3}{r^3}\right)$, so the magnitude of the slope increases as $r$ approaches $r_0$.

In our experiments, $r_0$ ranges between 5-13 µm, with a mean value of $r_0$=8 µm (Fig. S1). For a homogenously contracting network with $r_0$=8 µm, the non-linear correction term which scales like $\frac{r_0^3}{r^3}$ will shift the magnitude of the slope of a linear fit to the velocity profile in the range considered ($5\mu m < r - r_0 < 30\mu m$) by ~10%. The determination of the contraction rate from the slope of the experimental velocity profiles ignores this correction, which is comparable to the typical droplet-to-droplet variation in our data. Alternatively, we can calculate the volume contraction rate directly from the data assuming spherical symmetry by computing

$-div(V) = -\frac{1}{r^2}\frac{\partial}{\partial r}(r^2 V)$ (as in Fig. 3d).

We note that for values of $r$ close to $r_0$, the experimentally measured velocity profiles differ from the velocity profile of a homogenously contracting network. Specifically, the magnitude of the slope of the experimental velocity profiles close to $r_0$ decreases, indicating that the rate of network contraction decreases near the inner boundary. In these regions, where network density is high and network flow is slow, our assumption that the viscous stress dominates no longer holds and the elastic component of the stress likely slows down contraction.

# Supplementary Figures

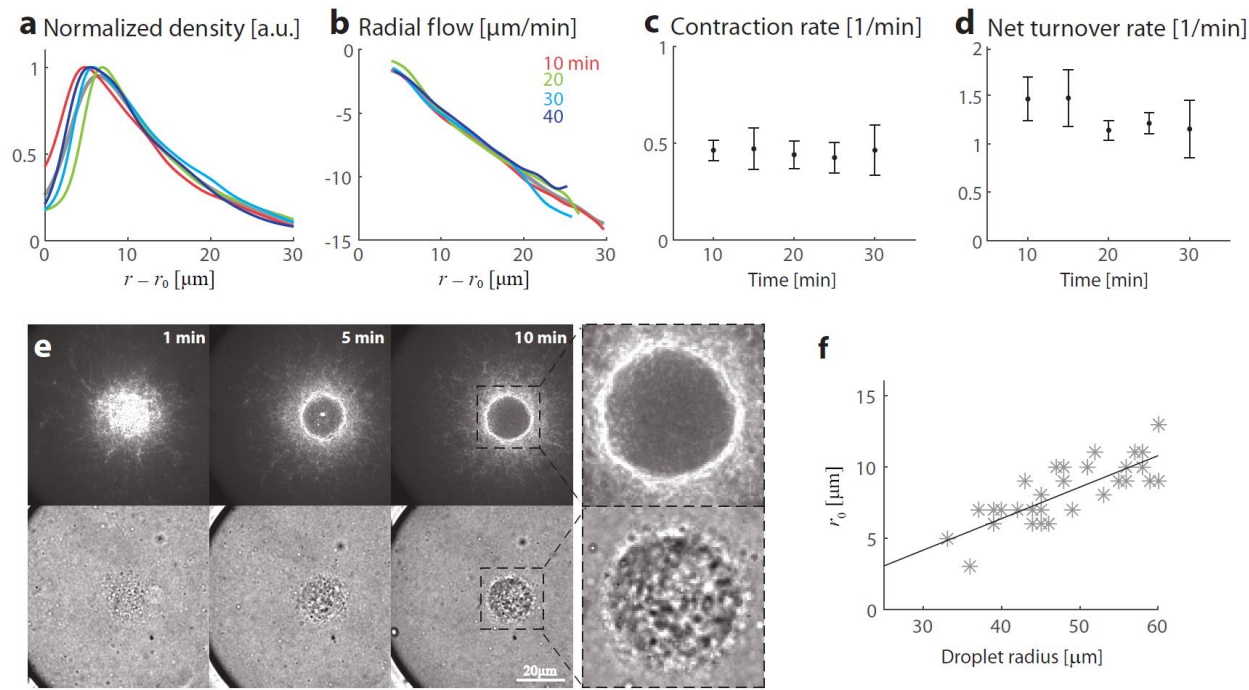

**Figure S1. Steady-state dynamics in confined contractile actomyosin networks.** (a-d) The network dynamics are measured in individual droplets as a function of time, at 5 min time intervals. (a,b) The normalized actin network density (a) and radial velocity (b) are plotted as a function of distance from the contraction center for one droplet at different time points. (c,d) The average contraction rate (c) and turnover rate (d) are depicted as a function of time (mean ± standard deviation; $N$=6). The contraction rate and net turnover rate are determined from the slopes of linear fits to the radial network flow as a function of distance, and the net turnover as a function of density, respectively. The characteristics of the system remain nearly unchanged, indicating that the system is at steady-state. (e) Bright-field (bottom) and spinning disk confocal images (top) from a time-lapse movie of the equatorial cross section of a droplet showing the formation of the exclusion zone around the contraction center in the first minutes after sample preparation. After 10 min the exclusion zone stops growing and the system reaches a steady-state. (f) The radius of the exclusion zone in different droplets is plotted as a function of the radius of the droplet.

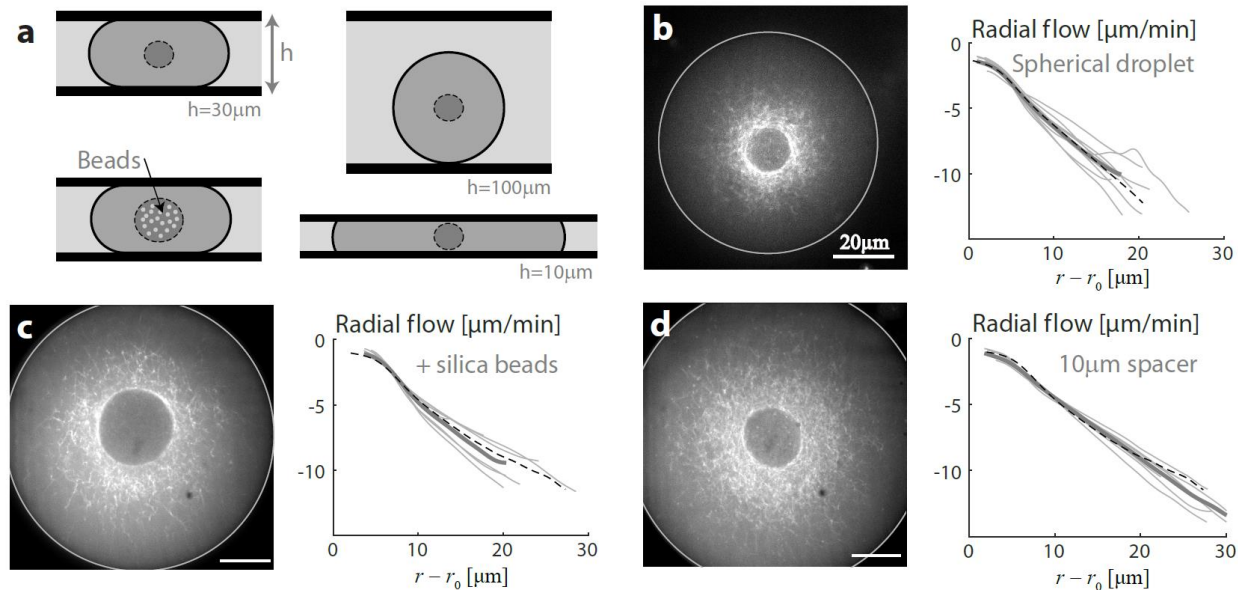

**Figure S2. Network contraction does not depend on sample geometry.** The contractile network behavior is characterized in different sample geometries. (a) Schematic illustrations of the different geometries: squished droplets in a 30 μm-height chamber (standard geometry); (b) Spherical droplets (imaged in a 100 μm-thick chamber); (c) Squished droplet with added beads that increase the size of the inner exclusion zone, and (d) Squished droplets in a 10 μm-height chamber. For (b) and (d) the samples were made normally, the only difference being the thickness of the sample chamber. For (c) the extract mix is supplemented with 1μm beads (which are pre-passivated by incubation with extract). The beads are swept inward by the contractile flow, and increase the diameter of the inner exclusion zone from ~20% of the droplets' diameter, to ~30% of the droplets' diameter. For each geometry, a spinning disk confocal image of the equatorial cross section of the network labeled with GFP-lifeact (left), is shown together with a graph of the radial network flow as a function of distance from the inner boundary (right). The thin grey lines depict data from individual droplets, and the thick lines depict the average over different droplets. The dashed lines show the results for the standard geometry (Figure 1e). In all cases, the radial velocity depends linearly on the distance from the inner boundary with the same slope, indicating that the networks exhibit homogenous contraction with the same contraction rate, irrespective of the sample geometry.

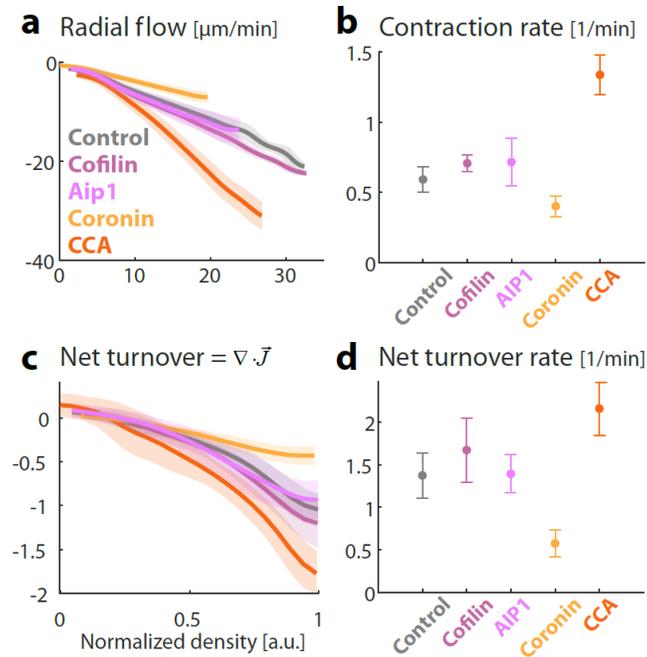

**Figure S3. Networks dynamics with added components of the actin disassembly machinery, Cofilin, Coronin and Aip1.** Contractile actin networks are generated in 'artificial cells' by encapsulating *Xenopus* extract supplemented with either 12.5µM Cofilin, 1.3µM Coronin, 2.6 µM Aip1 separately, or all three proteins together (CCA: 12.5µM Cofilin, 1.3 µM Coronin, and 1.3 µM Aip1). (a) The radial velocity is plotted as a function of distance from the inner boundary for the different conditions. For each condition, the mean (line) and standard deviation (shaded region) over different droplets are depicted. (b) The average contraction rate for each condition, determined from the slopes of linear fits to the radial velocity as a function of distance in individual droplets (mean ± standard deviation). (c) The net actin turnover as a function of the normalized network density is plotted for the different conditions. For each condition, the mean (line) and standard deviation (shaded region) over different droplets are depicted. (d) The average net turnover rate for each condition, determined from the slopes of linear fits of the net turnover as a function of network density. The net turnover rate and the contraction rate increase appreciably only when all three CCA proteins are added together. Addition of Coronin on its own, had an opposite effect. This may be due to Coronin's bundling activity [11], since similar effects are seen with the addition of the filament bundling protein Fascin (Fig. S5b).

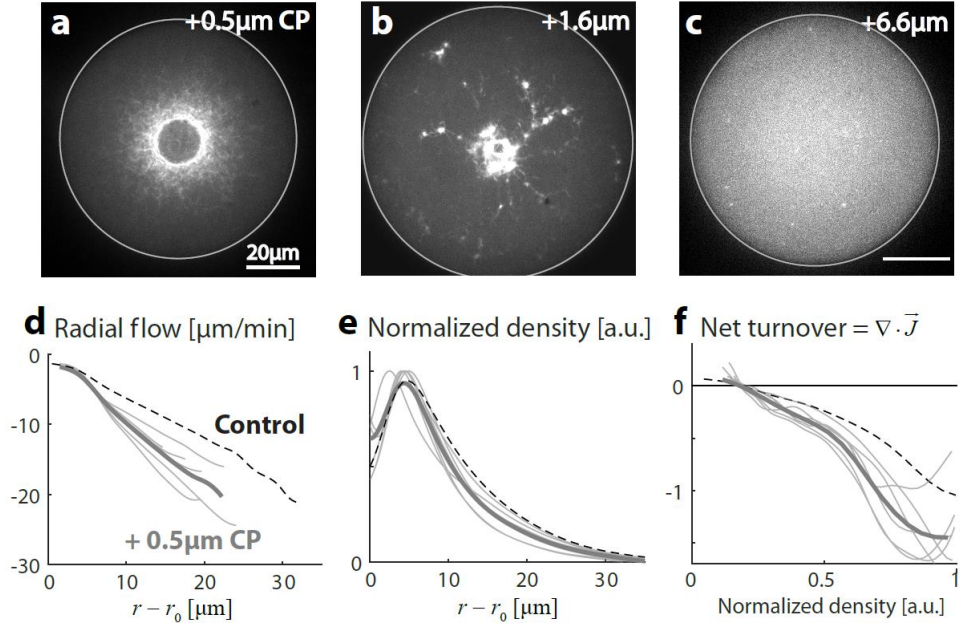

**Figure S4. Networks dynamics with added Capping Protein.** Contractile actin networks are generated in 'artificial cells' by encapsulating *Xenopus* extract supplemented with different concentrations of Capping Protein (CP), which caps barbed ends and blocks actin assembly there. (a-c) Spinning disk confocal fluorescence images of the equatorial cross sections of networks labeled with GFP-Lifeact in samples supplemented with 0.5, 1.6 or 6.6μM CP, respectively. (a) At a concentration of 0.5μM CP, the system reaches a steady-state within minutes, with a density distribution similar to unsupplemented samples. (b) At a concentration of 1.6μM CP, the network is composed of loosely connected clusters that contract towards each other. (c) At even higher CP concentrations, network formation is entirely blocked due to excess capping activity, and the probe density becomes homogenous. (d-e) The inward contractile flow and actin network density are measured for samples supplemented with 0.5μM CP. Graphs depicting the radial network flow (d) and density (e) as a function of distance from the inner boundary are shown. (f) Net actin turnover as a function of network density. The thin grey lines depict data from individual droplets, and the thick line is the average over different droplets. The dashed lines show the results for the control unsupplemented sample. Blocking assembly with CP leads to higher contraction rates and faster net turnover.

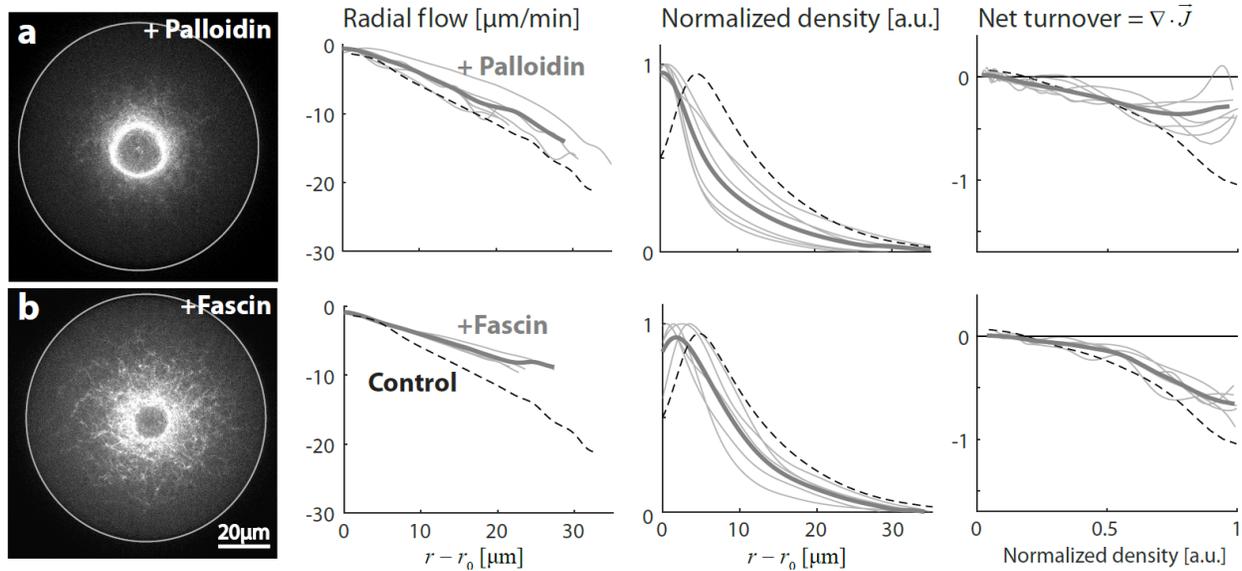

**Figure S5. Network dynamics under additional perturbations.** Contractile actin networks are generated in 'artificial cells' by encapsulating *Xenopus* extract supplemented with (a) 0.8μM Phalloidin which stabilizes actin filaments; or (b) 2.6μM Fascin which bundles actin filaments. In both cases, the system reaches a steady-state within minutes. The inward contractile flow and actin network density are measured (as in Figure 1). For each condition, a spinning disk confocal fluorescence image of the equatorial cross section of the network labeled with GFP-Lifeact (left) is shown, together with graphs depicting the steady-state radial network flow and density as a function of distance from the inner boundary (middle), and the net actin turnover as a function of network density (right). The thin grey lines depict data from individual droplets, and the thick line is the average over different droplets. The dashed lines show the results for the control unsupplemented sample. Stabilizing actin filaments with Phalloidin or adding the filament bundler Fascin leads to reduced turnover and slower contraction.
3434

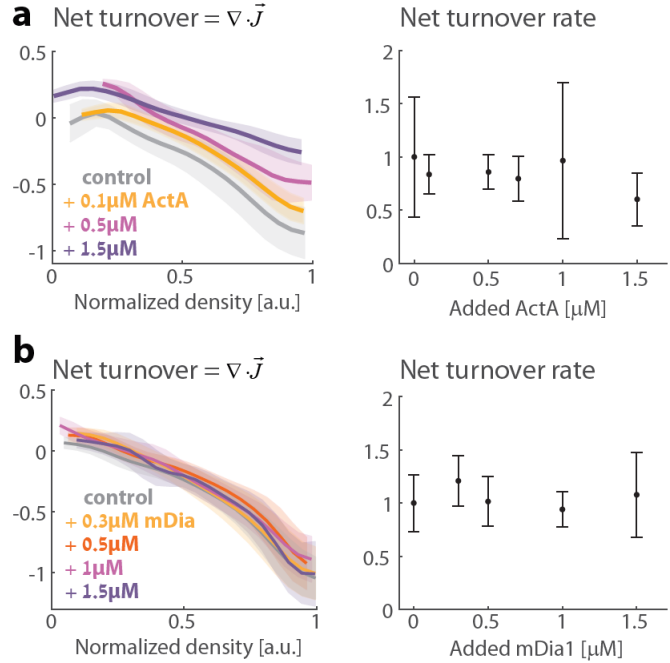

**Figure S6. Net network turnover with added assembly promoting factors.** Contractile actin networks are generated in 'artificial cells' by encapsulating *Xenopus* extract supplemented with different concentrations of ActA (a) or mDia1 (b). The divergence of the network flux, which is equal to the net network turnover, is plotted as a function of the local normalized network density (left). The net turnover rate (determined from the slope of a linear fit to the net turnover as a function of density) is plotted as a function of the added protein concentrations (right).

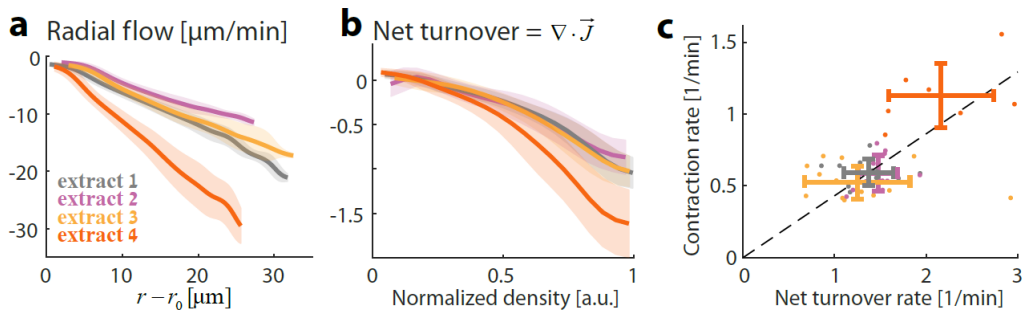

**Figure S7. Network dynamics for different extract batches.** Contractile actin networks are generated in 'artificial cells' by encapsulating different batches of *Xenopus* extract. (a) The radial network flow rate is plotted as a function of distance from the inner boundary. (b) The net actin turnover as a function of network density is plotted. For each batch of extract, the mean (line) and standard deviation (shaded region) over different droplets are depicted. The extract denoted 'extract 1' (grey) is the one used for most of the comparative data presented. (c) A scatter plot of the net contraction rate and net turnover rate for the different extract batches tested. The contraction rate and turnover rate are determined for each droplet from the slopes of linear fits to the radial network flow as a function of distance, and the net turnover as a function of density, respectively. For each extract, the dots depict values for individual droplets and the error bars show the mean and standard deviation of all the droplets examined for that extract.

## Supplementary Movies:

**Movie 1. Steady-state dynamics in a confined contractile actomyosin network.** This movie shows spinning disc confocal images at the equatorial plane of a contracting actomyosin network labeled with GFP-lifeact within an 'artificial cell'. The field of view is 105μm wide, and the elapsed time is indicated.

**Movie 2. Contractile actomyosin network supplemented with Cofilin, Coronin and Aip1.** This movie shows spinning disc confocal images at the equatorial plane of a contracting actin network labeled with GFP-lifeact formed with extract supplemented with 12.5μM Cofilin, 1.3μM Coronin and 1.3μM Aip1. The field of view is 105μm wide, and the elapsed time is indicated.

**Movie 3. Contractile actomyosin network supplemented with ActA.** This movie shows spinning disc confocal images at the equatorial plane of a contracting actin network labeled with GFP-lifeact formed with extract supplemented with 0.75μM ActA. The field of view is 105μm wide, and the elapsed time is indicated.

**Movie 4. Contractile actomyosin network supplemented with α–actinin.** This movie shows spinning disc confocal images at the equatorial plane of a contracting actin network labeled with GFP-lifeact formed with extract supplemented with 10μM α–actinin. The field of view is 105μm wide and the elapsed time is indicated.

**Supplementary References**

1. Bausch, A.R., F. Ziemann, A.A. Boulbitch, K. Jacobson, and E. Sackmann, *Local measurements of viscoelastic parameters of adherent cell surfaces by magnetic bead microrheometry.* Biophysical journal, 1998. **75**(4): p. 2038-2049.
2. Keller, M., R. Tharmann, M.A. Dichtl, A.R. Bausch, and E. Sackmann, *Slow filament dynamics and viscoelasticity in entangled and active actin networks.* Philosophical Transactions of the Royal Society of London A: Mathematical, Physical and Engineering Sciences, 2003. **361**(1805): p. 699-712.
3. Kole, T.P., Y. Tseng, I. Jiang, J.L. Katz, and D. Wirtz, *Intracellular mechanics of migrating fibroblasts.* Molecular biology of the cell, 2005. **16**(1): p. 328-338.
4. Wottawah, F., S. Schinkinger, B. Lincoln, R. Ananthakrishnan, M. Romeyke, J. Guck, and J. Käs, *Optical rheology of biological cells.* Physical review letters, 2005. **94**(9): p. 098103.
5. Panorchan, P., J.S.H. Lee, T.P. Kole, Y. Tseng, and D. Wirtz, *Microrheology and ROCK signaling of human endothelial cells embedded in a 3D matrix.* Biophysical journal, 2006. **91**(9): p. 3499-3507.
6. Rubinstein, B., M.F. Fournier, K. Jacobson, A.B. Verkhovsky, and A. Mogilner, *Actin-myosin viscoelastic flow in the keratocyte lamellipod.* Biophysical journal, 2009. **97**(7): p. 1853-1863.
7. Zhu, C. and R. Skalak, *A continuum model of protrusion of pseudopod in leukocytes.* Biophysical journal, 1988. **54**(6): p. 1115-1137.
8. Jin, T., L. Li, R.C.M. Siow, and K.-K. Liu, *A novel collagen gel-based measurement technique for quantitation of cell contraction force.* Journal of The Royal Society Interface, 2015. **12**(106): p. 20141365.
9. Yang, M.T., D.H. Reich, and C.S. Chen, *Measurement and analysis of traction force dynamics in response to vasoactive agonists.* Integrative Biology, 2011. **3**(6): p. 663-674.
10. Yamada, S., D. Wirtz, and S.C. Kuo, *Mechanics of living cells measured by laser tracking microrheology.* Biophysical journal, 2000. **78**(4): p. 1736-1747.
11. Goode, B.L., J.J. Wong, A.C. Butty, M. Peter, A.L. McCormack, J.R. Yates, D.G. Drubin, and G. Barnes, *Coronin promotes the rapid assembly and cross-linking of actin filaments and may link the actin and microtubule cytoskeletons in yeast.* J Cell Biol, 1999. **144**(1): p. 83-98.